\documentclass[twocolumn,superscriptaddress,floatfix,prl,aps,showpacs]{revtex4}
\usepackage{graphicx}
\begin{document}
\bibliographystyle{apsrev}
\title{N\'eel order in doped quasi one-dimensional antiferromagnets}
\author{Sebastian Eggert}
\affiliation{Institute of Theoretical Physics,
Chalmers University of Technology and G\"oteborg University,
S-412 96 G\"oteborg, Sweden}
\author{Ian Affleck}
\affiliation{Physics Department, Boston University,
 Boston, MA 02215}
\altaffiliation[On leave from ]{Canadian Institute for 
Advanced Research and Department of Physics
 and Astronomy, University of British Columbia, Vancouver,
BC,  Canada, V6T 1Z1}
\author{Matthew D.P. Horton}
\affiliation{99 John Street, New York, NY 10038}
\date{March 29, 2002. Last change: \today}
\begin{abstract}
We study the N\'eel temperature
of quasi one-dimensional S=1/2 antiferromagnets containing non-magnetic 
impurities.  We first consider the temperature dependence of the 
staggered susceptibility of finite chains with open boundary conditions, 
which shows an interesting difference for even and odd length chains.
We then use a mean field theory treatment to incorporate the three dimensional 
inter-chain couplings.  The resulting N\'eel temperature
shows a pronounced drop as a function of doping by up to a factor of 5.
\end{abstract}
\maketitle
The study of doped low-dimensional antiferromagnets has been a very active
field since the discovery of high-T$_c$ superconductivity.  A particularly 
simple form of doping results from replacing some magnetic Cu ions by  non-magnetic 
ions like Zn.  In this case the system is well described by the Heisenberg
model with some spins removed from a regular lattice. This problem has been 
quite extensively studied both theoretically and experimentally in 
the quasi-two-dimensional case.  Recent Monte Carlo simulations have shown 
convincingly that the zero temperature 2 dimensional system remains N\'eel
ordered for impurity concentrations, $p$ 
up to the classical percolation threshold, ($p_c\approx .407$, 
on the square lattice) \cite{Kato,Sandvik}. An 
analytic approach has been developed based on spin-wave theory and the
T-matrix approximation, valid at impurity concentrations well below
percolation, and extended to include weak inter-plane couplings \cite{Chernyshev}.
Recent experiments on La$_2$Cu$_{1-x}$(Zn,Mg)$_x$O$_4$ \cite{Vajik} are 
consistent with the $T=0$ critical concentration corresponding to classical 
percolation, and agree in detail, at lower impurity concentrations, with 
the analytic theory.  

 Here we study the 
effect of doping on three dimensional ordering in spin-1/2 chain compounds where
the exchange interaction along the chains $J$  is much stronger than the 
inter-chain coupling $J'$. Before doping, the Hamiltonian is
\begin{equation}
H =  \sum_{j,\vec{y}} \left(J \vec{S}_{j,\vec{y}} \cdot \vec{S}_{j+1,\vec{y}} 
+ \sum_{\vec{\delta}} J' \vec{S}_{j,\vec{y}} \cdot \vec{S}_{j,\vec{y}+\vec{\delta}}\right),
\label{spinham}
\end{equation}
where $J$ is the site-index along the chains and 
$\vec{\delta}$ are the vectors to neighboring chains.  
The chain lattice spacing has been set to unity.
 Randomly removing some of the spins breaks the chains up
into finite segments with open boundary conditions, which  are still weakly coupled
to neighboring chains.  This model describes  Zn doped
Sr$_2$CuO$_3$, for example.  For the pure system a standard method to study
the N\'eel temperature for weakly coupled chains is to first determine the 
staggered susceptibility of the one-dimensional chains, $\chi_1(T)$. 
If we then treat the inter-chain couplings 
in mean field theory \cite{MFT} we obtain the condition which determines the N\'eel
temperature
\begin{equation} zJ'\chi_1(T_N)=1,\label{MFT}\end{equation}
where $z$ is the number of neighboring chains from the sum over 
$\vec{\delta}$ in Eq.~(\ref{spinham}).  Since at low
$T$, $\chi_1 (T)$ diverges as $1/T$, this predicts $T_N\propto J'$, so that
N\'eel order is predicted to occur for arbitrarily weak inter-chain coupling.  

In this paper we extend this approach to the doped system 
by calculating the staggered susceptibility of chains with arbitrary length $L$
to find an average value of $\chi_1$ as a function of temperature $T$.
 This type of mean field treatment of {\it inter-plane} couplings in 
doped samples was used to study  the N\'eel 
temperature of quasi-two-dimensional antiferromagnets in \cite{Chernyshev}.  
This method can be carried out much more accurately in the
quasi-one-dimensional case studied here because an analytic expression 
for the staggered 
susceptibility of finite chains can be found, which 
by itself yields rather interesting results, exhibiting very different behavior
for even and odd length chains. Eq.~(\ref{MFT}) corresponds to approximating 
the interchain interactions as simply providing a staggered field of fixed 
magnitude, acting on a given chain.  This approximation results from 
averaging over both quantum fluctuations and impurity locations on 
neighboring chains.  We expect it to become more reliable when 
the average chain length $\bar L$ and $z$ (i.e.~the lattice dimension) increases.  
 Clearly this approximation must break down 
at large impurity doping $p =1/\bar L$, before the percolation threshold is reached
($p_c\approx 69.8\%$ for a three dimensional simple cubic lattice).  In lower
dimensions this approach becomes more questionable since the percolation 
threshold is reached earlier and the number of neighbors is lower, while
in higher dimensions this method may become exact as $z\to \infty$ and valid
for all doping levels since $p_c \to 1$. 

The staggered susceptibility per unit length 
of a finite chain of length $L$ with 
open boundary conditions at finite temperature
$T = 1/\beta$ is
\begin{equation}
\chi_1 (L,T)\equiv \frac{1}{L}\int_0^\beta d\tau \sum_{j,k=1}^L (-1)^{j-k} \langle
S^z_j(\tau) S^z_k(0) \rangle .
\label{chialt}
\end{equation}
In order to find a reliable average we need
to determine $\chi_1$ for a large range of temperatures and lengths for which we 
use both bosonization techniques and numerical Monte Carlo simulations.
Note that here we measure the staggered response to a staggered 
field, not to be confused with the staggered response to a uniform 
field \cite{impurity}.

In order to calculate $\chi_1$ as a function of temperature we now go to the 
continuum limit and use the field theory treatment which is correct 
in the asymptotic low $T$, large $L$ limit \cite{review}.  The spin operators are then 
described in terms of a boson field, $\phi$, 
\begin{equation}
S^z_j \approx \frac{\partial_x \phi}{\sqrt{2 \pi}} + C (-1)^j \cos \sqrt{2 \pi} \phi,
\label{sz}
\end{equation}
where $C$ is a  parameter which will be discussed in more detail below.
The boson field $\phi$ is described by a free massless relativistic Hamiltonian \cite{review}
 up to a marginally irrelevant interaction which gives rise to logarithmic corrections 
as we will see later.
We normalize the operator $\cos \sqrt{2\pi}\phi $ so that its $T=0$,
$L=\infty$, equal time correlation function decays with distance as $1/2|r|$.
In the sum of Eq.~(\ref{chialt}) we can neglect all rapidly oscillating parts so that
only the second term in Eq.~(\ref{sz}) will be kept in the alternating 
spin-spin correlation function
\begin{equation}
G(x,y,\tau) = C^2 \langle \cos \sqrt{2 \pi} \phi(x,i \tau) \cos\sqrt{2 \pi} \phi(y,0) \rangle,
\end{equation}
so that Eq.~(\ref{chialt}) becomes
\begin{equation}
\chi_1 \approx  \frac{1}{L}\int_0^\beta d\tau \int_0^L dx \int_0^L 
dy \, G(x,y,\tau).
\label{integral}
\end{equation}
To calculate the correlation function we use the mode expansion 
of the boson for a finite chain with open boundary conditions \cite{ourPRB}
\begin{equation}
\phi(x,t) = \sqrt{\frac{\pi}{8}} + \sqrt{2 \pi} S^z \frac{x}{L} + \phi_>(x,t),\label{exp}
\end{equation}
where
\begin{equation}
\phi_>(x,t) \equiv \sum_{n = 1}^\infty 
\frac{1}{\sqrt{\pi n}} \sin \frac{\pi n x}{L}\left(e^{-i \pi n v t/L} a_n +
  h.c. \right)
\end{equation}
contains the ordinary boson modes $a_n$ and 
the ``zero mode'' eigenvalue $S^z$  corresponds to the z-component of the total
spin of the chain, which takes integer values for even lengths $L$ and
half-integer values for odd $L$.  The spin wave velocity is given by $v= \pi J/2$.

Before considering the case of arbitrary $T$ and $L$, it is interesting to 
consider the limit $T\to 0$ with $L$ held fixed. In this limit, upon 
inserting a complete set of states,  
$\chi_1 \to {1\over LT}|\langle 0|S^z_{\rm alt}|0\rangle |^2$ will be dominated by the 
groundstate $|0\rangle$, where 
$S^z_{\rm alt}\equiv \sum_{j=1}^L(-1)^{j+1}S^z_j$.
 Using Eq.~(\ref{exp}) we can directly find the 
{\it local} staggered magnetization of the lowest energy state in any sector with a given $S^z$ 
\begin{eqnarray}
\langle S^z(x)\rangle & \approx &  C (-1)^x\langle \cos \sqrt{2\pi}\phi (x)\rangle \nonumber \\
& = & C(-1)^x\frac{\sin (2\pi S^zx/L)}{\sqrt{(2 L/\pi) \sin (\pi x/L)}}.
\end{eqnarray}
For $L$ even $S^z=0$ in the groundstate so this gives zero $L T \chi_1  \to 0$.  However, for 
$L$ odd $S^z=\pm 1/2$ in the groundstates, giving
\begin{equation}
\langle\pm |S^z(x)|\pm \rangle \approx  \pm C (-1)^x\sqrt{\frac{\pi}{2L} \sin \frac{\pi
  x}{L}}
,\label{Szj}\end{equation}
which upon integrating over $x$ gives $S^z_{\rm alt} \propto \sqrt{L}$ and 
therefore $\chi_1 \propto 1/T$. This divergence is in sharp contrast to the even case.     
Interestingly Eq.~(\ref{Szj}) indicates a maximum response in the center of the chain
which agrees with numerical results \cite{laukamp,long} and is reminiscent of
the square-root increase of the staggered response to a uniform field 
with the distance from the open ends \cite{impurity}.
It is interesting to note that our finding  $S^z_{\rm alt} \propto \sqrt{L}$ for a spin chain
corresponds to an intermediate result between a N\'eel state with $S^z_{\rm alt}=\pm L/2$
and a  nearest neighbor dimer state with one unpaired spin 
$S^z_{\rm alt}=\pm 1/2$.  

We now consider the case of general $L$ and $T$ using the field theory approach.  
The correlation function can then be written as
\begin{eqnarray}
& & G(x,y,\tau) =     \frac{C^2}{2}\left(
   \langle e^{ i 2 \pi S^z (x-y)/L}\rangle \langle e^{i \sqrt{2 \pi}(\phi_>(x,i \tau)
- \phi_>(y,0))}\rangle \right. \nonumber \\ 
 & &  \ \ \ \ \ -  \left. \langle e^{ i 2 \pi S^z (x+y)/L}\rangle
\langle e^{i \sqrt{2 \pi}(\phi_>(x,i \tau) + \phi_>(y,0))}\rangle \right).
\end{eqnarray}
Upon using the cumulant theorem for boson 
modes $\langle e^A\rangle  = e^{\langle A^2\rangle/2}$
we can determine the correlation function at any finite temperature and length by following 
the analogous calculations in Refs.~\cite{mattsson} and
\cite{EMK}.  Using the shorthand notation $u = \pi \frac{x-y-i v\tau}{2 L} $,  
$\bar{u} = \pi \frac{x-y+iv \tau}{2 L} $, $w = \pi \frac{x+y+i v\tau}{2 L} $, and $\bar{w}
 = \pi \frac{x+y-i v \tau}{2 L} $ we find 
\begin{eqnarray}
 & &  G(x,y,\tau)   =   
\frac{\pi C^2}{4 L} \frac{\partial_x \theta_1(0, e^{-\gamma})}{\sqrt{\theta_1(w+u, 
e^{-\gamma})\, \theta_1(w-\bar{u}, e^{-\gamma})}} \nonumber \\
& &  \ \ \times
 \left[ B\left( u+\bar{u},e^{-2\gamma}\right) 
\sqrt{\frac{\theta_1(w, e^{-\gamma})\theta_1 ( \bar{w}
, e^{-\gamma})}{\theta_1(u, e^{-\gamma}) 
\theta_1(\bar{u} , e^{-\gamma})}}\right. 
\label{coscos} \\ 
  & &\ \ \   -\ \left.B\left( w + \bar{w},e^{-2\gamma}\right) 
\sqrt{\frac{\theta_1(u, e^{-\gamma})\theta_1 (\bar{u} 
, e^{-\gamma})}{\theta_1(w, e^{-\gamma}) 
\theta_1(\bar{w}, e^{-\gamma})}}\right], \nonumber
\end{eqnarray} 
where 
$\theta_1$ is the elliptic theta function of the first kind \cite{GR}.   
The parameter $\gamma= \frac{v \pi}{2 L T}$
gives the spacing in the finite size energy spectrum  in relation to 
the temperature. The contribution $B(x)$ from the zero modes is given by
\begin{equation}
B(z,e^{-2\gamma}) \equiv {\sum_{S^z}e^{-2\gamma (S^z)^2+2
iS^zz}
\over \sum_{S^z}e^{-2\gamma (S^z)^2}}=
\frac{\theta (z, e^{-2 \gamma)}}{\theta(0,e^{-2 \gamma})},
\end{equation}
where $\theta$ is the elliptic theta function of the second kind
for odd chains $\theta = \theta_2$, while it is the elliptic theta function of the 
third kind for even chains $\theta = \theta_3$. 
Remarkably, the correlation functions in 
 the continuum limit in Eq.~(\ref{coscos}) therefore retain information about
the underlying lattice and explicitly depend on the parity of $L$. 
This result  requires the explicit use of the zero modes in the 
mode expansion \cite{mattsson,EMK}.  The difference arises 
because of the different set of eigenvalues of $S^z$: integer and half-integer
for even and odd length chains, respectively.

At this point we may rescale all the variables of integration in 
Eq.~(\ref{integral}) by $L$ (or alternatively $T$)
to express $\chi_1$ in terms of a universal function of the dimensionless 
variable $LT/v$.
\begin{equation}
\chi_1 = C^2 f(LT/v)/T  \label{scaling}
\end{equation}
\begin{figure}
\includegraphics[width=.48\textwidth]{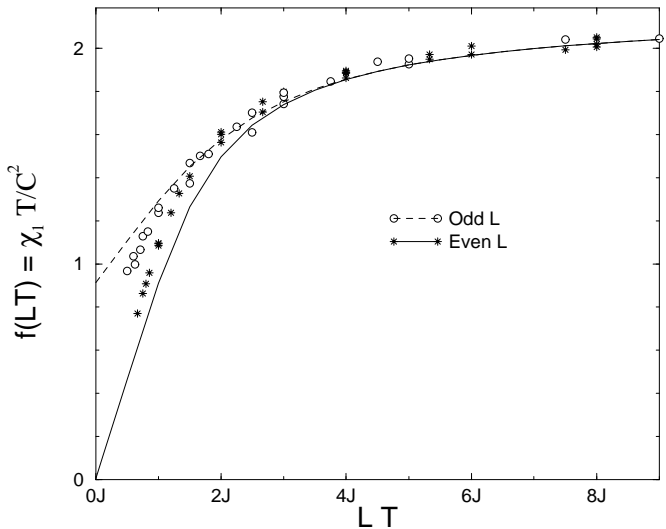}
\caption{The scaled staggered susceptibility in Eq.~(\ref{scaling}) 
$f(LT) = \chi_1 T/C^2$ determined from 
Eqs.~(\ref{integral}) and (\ref{coscos}) 
for even  and odd  length chains. The upper boundary of the 
graph represents the asymptotic limit as $L T \to \infty$ in Eq.~(\ref{finiteT}).  
The points correspond to numerical 
Monte Carlo results for different combinations of $L$ and $T$ divided by the logarithmic 
correction in Eq.~(\ref{logs}) with $a=23$. }
\label{chialtfig}
\end{figure}
In the thermodynamic limit $LT/v\to \infty$, we can use 
the asymptotic behavior of the $\theta$ functions \cite{GR} as was done 
in a related calculation in \cite{mattsson,EMK} for $e^{-\gamma} \to 1$.
In this case we can combine the two terms of $G(x,y,\tau)$ in Eq.~(\ref{coscos})
into one,
giving the known finite temperature correlation functions \cite{impurity}.
This results in the well-known $1/T$ behavior \cite{benzoate} 
\begin{equation}
\chi_1
 \ \stackrel{\gamma \to 0}{\longrightarrow}\    \frac{C^2}{4T}
\frac{\Gamma^2(1/4)}{\Gamma^2(3/4)} 
\ \approx \ 2.18844 \ C^2  / T
\label{finiteT}
\end{equation}
As expected there is no difference between even and odd length chains in the limit of
$L T/v \to \infty$ in Eq.~(\ref{finiteT}).

In the opposite limit of zero temperature and finite length  $LT/v \to 0$, however, 
we find a qualitative difference for even and odd chains.  
Again using asymptotic limits of $\theta$ functions as $e^{-\gamma} \to 0$, we now find 
\begin{widetext}
\begin{equation}
G(x,y,\tau) \to \frac{\pi C^2}{4L}
\sqrt{\frac{\sin(w+u) \sin(w-\bar{u})}{\sin(u) \sin(\bar{u}) \sin(w)
\sin(\bar{w})}}
\times \left\{
\begin{array}{lcl}
 1, & \phantom{nn}& \text{L  even} \\
 & & \\
\cosh(\frac{\pi v\tau}{L}), & \phantom{nn}& \text{L  odd} \\
\end{array} \right.
\label{zeroT}
\end{equation}
\end{widetext}
Since for odd $L$ the correlation function $G(x,y,\tau)$ approaches a constant 
as $\tau \to \infty$ we get a divergence of $\chi_1$ in Eq.~(\ref{integral}) with
$1/T$ for low temperatures, while for even $L$ the integral is proportional to $L$
resulting in
\begin{equation}
\chi_1   \
\stackrel{\gamma \to \infty}{\longrightarrow} \
\left\{
\begin{array}{lcl}
 0.92905 \ C^2 L/J,&\phantom{n}&\text{L even}\\
&~ & \\
 \frac{8 C^2}{ \pi T} E^2\left( \frac{\pi}{4},\sqrt{2}\right)
\approx \ 0.913893\  C^2/T,
& \phantom{n}& \text{L  odd}
\end{array}
\right. \label{LTE}
\end{equation}
where $E$ is the elliptic integral of the second kind \cite{GR}, which can also be derived 
from Eq.~(\ref{Szj}).  Note that the scaling behavior with $1/T$ in the two 
limits $LT/v\ll 1$ for odd $L$
in Eq.~(\ref{LTE}) and $LT/v\gg 1$ in Eq.~(\ref{finiteT}) is 
the same, up to a factor of about 2. 
 This is of crucial significance for the behavior of the 
N\'eel temperature of the doped quasi-one-dimensional system 
as we shall see. 

So far we have ignored the marginally irrelevant interaction  mentioned
earlier.  Its effect on the staggered susceptibility at finite $T$ but 
$L\to \infty$ is well-known.  It corresponds to replacing the constant $C^2$
by a slowly varying function of $T$
\begin{equation}
C^2 = 2 \sqrt{\ln(aJ/T)/(2 \pi)^3}, 
\label{logs}
\end{equation}
where $a$ is a dimensionless constant \cite{exact,benzoate}. 
From fitting our susceptibility data, we find $a\sim 23$, which gives 
$C^2 \approx 0.33 \pm 0.07$  for $0.001 \alt T/J \alt 0.2$. 
A finite length $L$ {\it together with 
open boundary conditions} leads to more complicated logarithmic corrections
which may in general involve a different exponent 
near the boundary \cite{Qin}.
Hence, the general expression of the logarithmic correction at finite $L$ is 
not known, but we expect that $C^2 \approx 0.33 \pm 0.07$ remains 
approximately correct for the relevant 
length scales studied here.  

The full behavior of $\chi_1$ as a function of the scaling variable $L T$
 is shown in Fig.~(\ref{chialtfig}) compared to numerical 
Monte Carlo data after dividing by the logarithmic factor  
in Eq.~(\ref{logs}). 
There are no adjustable parameters for this fitting except 
for the constant inside the logarithm $a\approx 23$ and all numerical points from 
the Monte Carlo simulations fall close to this universal line for larger values of 
$L T$  as well (not shown). The errors are less than the size of the symbols in the 
figure so that the deviations are due to higher order corrections.  
For the simulations we chose
different values of $20 \alt L \alt 120$ and $0.025 \alt T/J \alt 0.2$.
For $L T \agt 4J$ the even and odd cases are virtually indistinguishable, but 
as $L T \to 0$ there is a clear difference in the behavior. 
\begin{figure}
\includegraphics[width=.48\textwidth]{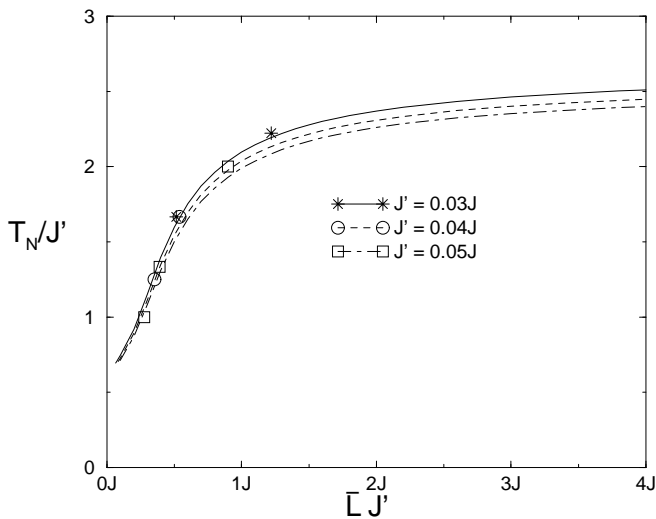}
\caption{The N\'eel temperature $T_N/J'$ as a function of $J'$ and average chains length $\bar{L}$ from bosonization results (lines) compared to Monte Carlo simulations (symbols).}\label{TN}
\end{figure}

We now want to determine the N\'eel temperature by using the mean field
treatment of inter-chain couplings in Eq. (\ref{MFT}), where $\chi_1$ now 
represents the one-dimensional susceptibility averaged over chain lengths with 
the probability distribution, $P(L)= e^{-L/\bar L}/\bar L$, where $\bar L$ is 
the average chain length, corresponding to an impurity concentration  of $p=1/\bar{L}$.
Because of the scaling form in Eq.~(\ref{scaling}) it is straightforward to show that
the mean field condition in Eq.~(\ref{MFT}) can be written as
\begin{equation}
T_N = C^2 z J' g(T_N \bar{L}) \label{scaledMFT}
\end{equation}
where $g(T_N \bar{L}) = \int dy \ y e^{-y} f(T_N\bar{L}\, y)$ is the
average of the scaling function $f$ in  Fig.~(\ref{chialtfig}) and
Eq.~(\ref{scaling}) and $C^2$ is given in Eq.~(\ref{logs}) with $T=T_N$.
We see that the solutions for the N\'eel temperature $T_N/J'$ in  Eq.~(\ref{scaledMFT}) 
are functions of the scaling variable $\bar{L}J'$ as shown in Fig.~(\ref{TN}) for $z=4$
compared to the results from Monte Carlo simulations.  The marginal operator 
leads to weak logarithmic corrections to this scaling behavior,  which leaves the shape
of the curve largely unchanged for different $J'$, and only shifts it up by a few percent as
$J'$ is lowered.  Therefore we can make a nearly universal {\it quantitative} 
prediction for all doping levels and coupling strengths.

The N\'eel temperature is strongly affected by  doping when 
the impurity concentration, $1/\bar L\geq J'/J$ 
and may drop by as much as a factor of 5, although it remains finite.  This is
because the scaled average staggered susceptibility $g(T_N \bar{L})$ 
of odd chains is finite as $\bar{L} \to 0$ so that Eq.~(\ref{scaledMFT}) 
can always be fulfilled for a positive $T_N$.  If,  however, only 
even chains were allowed in the system, no non-zero solution would exist for 
$\bar{L} \alt 0.6J/z J'$. 
As mentioned above, we expect this result to break down at larger impurity 
doping as the percolation threshold is approached and N\'eel order
disappears.   

\begin{acknowledgments}
I.A. would like to thank Antonio Castro-Neto and Anders Sandvik for very
helpful conversations and Kenji Kojima for interesting him in this subject.  
This research was supported in part by the Swedish Research Council (SE) and
NSERC of Canada (IA and MDPH).
\end{acknowledgments}

\end{document}